\documentclass[%
preprint,
 amsmath,amssymb,
 aps,
 prc
]{revtex4-2}

\usepackage{graphicx}
\usepackage{dcolumn}
\usepackage{bm}

\begin{document}

\title{Nuclear EoS with finite range interaction and the Constrained Dynamics}

\author{Massimo Papa}
 \email{Massimo.Papa@ct.infn.it}
\affiliation{%
 Istituto Nazionale Fisica Nucleare-Sezione di Catania, Corso Italia 57, I-95129 Catania, Italy\\
}%


%

\date{\today}

\begin{abstract}
The effects of the many-body correlations  with respect the MF description characterizing the Constrained  Molecular Dynamics are discussed in the case of finite and zero range effective microscopic interactions. In particular, for the system $^{64}Ni+^{48}Ca$ fusion-fission cross-sections, IMF production and transversal flow have been evaluated
in the framework of the CoMD model at different incident energies.
The results obtained significantly depend on whether the effects of the above-mentioned correlations are taken into account in the corresponding EoS.
\end{abstract}

\maketitle
\section{Introduction}
The description of many-body systems is one of the most
difficult problems in nuclear physics due to their complexity being quantum objects described by
a large number of degrees of freedom.
In particular, the heavy ion collisions (HIC) at energy well above 
the mutual Coulomb barriers are usually described through semi-classical
approaches  based on the mean-field (MF) approximation or 
quantum molecular dynamics approaches (QMD) ~\cite{wolt}. These last describe the single
particle wave functions by means of well localized 
wave-packets (WPs) with fixed widths.  
In this way many-body correlations  are produced which  lead to the spontaneous formation of clusters.
In these semi-classical approaches the effective interaction 
plays obviously  a key role, and in many cases it just represents the main subject of investigation. 
In several cases MF and molecular dynamics approaches ~\cite{wolt} share the same  microscopic effective interactions. 
From a general point of view, it can be  expected
that the typical and explicit two or many-body correlations of QMD like 
approaches could instead play a  role in many-body quantities  as for the total energy.
At what extent these specific correlations may affect the many-body functional (EDF) related to the total energy in the case of infinite nuclear matter(NM) was discussed in some detail in Ref.~\cite{npa}
~\cite{comd}.
In particular, in this contribution, we aim to illustrate how these correlations influence some common observables studied in Heavy Ion Collisions (HIC) at the Fermi energies , such as fusion-fission and Intermediate Mass Fragment (IMF)production.

\noindent
In the following section the topic is introduced by specifying the microscpic effective interaction used and briefly discussing the procedure adopted to obtain the
set of parameters describing a reference Equation of State (EoS)
around the saturation density.  
In the other sections proceeding to finite systems the  results of calculation concerning
the $^{64}Ni+^{48}Ca$ collision at different energies are shown. 
In particular, for the case of finite and zero range
interaction (ZR) the comparison of different reaction mechanisms and the nucleon transverse flow for central/semi-central collisions is illustrated.
To understand at what extent the correlations produced by the model can affect the above observables, comparisons using the set of  parameter values obtained in the MF scheme are also shown.

\section{The Microscopic Effective interaction}
\noindent
In the following we will refer to an example of EoS characterized by some properties at the saturation density at zero temperature that we choose as reference. Equilibrium density 
$\rho_{0}=0.165$ $fm^{-3}$, associated 
binding energy
$E(\rho_{0})=-16$ MeV,  incompressibility $K(\rho_{0})=240$ MeV,  
symmetry energy $E_{sym}=30$ MeV, an effective pairing energy 
(it include the spin-dependent one produced trough the exchange terms \cite{npa})
equal to -2 MeV per nucleon at the saturation density in finite systems (around mass 100).
Moreover, the functional will correspond in the MF limit to a
relative effective mass $m^{*}_{r}=0.67$, 
and  neutron-proton
effective mass splitting $m^{*}_{rn}-m^{*}_{rp}=0.4\beta$. 
 Different values of the slope parameter associated to the symmetry energy $L=3\rho_{0}(\frac{dE_{sym}}{d\rho})_{\rho_{0}}$
have been considered changing in the range $ 55 \div 105$ MeV as suggested from different investigations 
(see as an example ~\cite{prcconstr}).
 The total microscopic interaction, inspired from the Gogny interaction,  will be the sum of two and three body terms with
 zero and finite range $\mu=1.1$ fm. The finite range is modelled according to Gaussian factors. 
Finally, a further term corresponding to  Eq.(1e) in \cite{npa}  represents a zero range spin-spin interaction. As explained  in the following section, this contribution
is necessary to reproduce  an effective "pairing" energy of about -2 MeV
at the ground state for finite system with mass around 100.
(see also the following section). 
At the same time,in box calculations, this further contribution is able to locally produce small values of the average total spin
at the stationary conditions.
The introduction of this term produce a further contribution to the
symmetry energy. 
\section{The Effective interaction}
Starting from  the microscopic interaction, the expression of the effective interaction as a function of the density can be obtained in the MF approximation
by evaluating the associated matrix elements using 2-body wave functions constructed with plane-waves in a large volume V. $\Phi=\frac{1}{\sqrt{V}}e^{i\textbf{kr}}$.
The zero range interactions will give arise to the usual terms proportional to powers of the density. The finite range interaction will give arise to a momentum dependent part because of   the underlying Slater determinant structure of the 2-body wave function for the identical particles. 
\textit{Contrary to the Gogny case, it has been supposed, according to the several experimental evidences collected on the Isospin symmetry breaking in nuclei
\cite{symmetry},  that the neutron and proton are different kinds of particles. Therefore in the present approach neutron-proton couples do not generate
exchange terms and the related momentum dependent interaction}.
This condition, as noted in the previous paragraph, required the introduction of the term associated with Eq.(1e) of Ref.  \cite{npa} in order to produce the desired effective mass and effective "pairing".
In the following as an example we write the two-body exchange contribution 
 \begin{equation}
\Delta E^{ex}_{2}(\textbf{k}_{i},\textbf{k}_{j})=-\frac{P_{2}}{V}(\sqrt{\pi}\mu)^{3}\textit{e}^{-\mu^{2}(\textbf{k}_{i}-\textbf{k}_{j})^{2}/4}
\end{equation}
The non-locality of the effective interactions produces 
corrective factors to the in medium nucleon kinetic energy formally represented through a density dependent nucleon effective mass $m^{*}$.
In the case of the CoMD model the first step is just to substitute the plane-waves convolution with the wave packets $\Phi=\frac{1}{(2\pi \sigma_{r}^{2})^{3/4}}
\textit{e}^{-\frac{(\textbf{r}-\textbf{r}_{0,i})^{2}}{4\sigma_{r}^{2}}+i\textbf{k}_{0,i}\textbf{r}}$.
As an example, the corresponding momentum dependent (MDI) 2-body contribution  will be:
\begin{equation}
E^{i,j,MDI}_{2}=-\frac{P_{2}}{8\sigma_{r}^{3}}\xi^{3}
\times\textit{e}^{-\frac{1}{4}[\frac{(\textbf{r}_{0,i}-\textbf{r}_{0,j})^{2}}{\sigma_{r}^{2}}+\xi^{2}
(\textbf{k}_{1}-\textbf{k}_{1})^{2}]} 
(\delta_{\tau_{i}-\tau_{j}}\delta_{s_{i}-s_{j}})
\end{equation}
With a modified width :$\frac{1}{\xi^{2}}=\frac{1}{4\sigma_{r}^{2}}+\frac{1}{\mu^{2}}$.
While the gaussian width of the related  direct contribution 
will be: $\lambda^{2}=4\sigma_{r}^{2}+\mu^{2}$.
The differences in the arguments of the exponentials appearing in equations (1) and (2) should be noted. They  are the principal reasons together with the Pauli constraint for the existence of typical correlations
in the molecular dynamics approach not present in the MF one \cite{npa} .

\noindent 
In the framework of the MF approximation, 
all the quantities characterizing the reference  functional 
can be obtained in a relatively simple way by
solving a linear system with the strength parameters as unknown quantities.
In the case of CoMD calculations the nuclear matter is simulated by means of box calculations with periodic boundary conditions.
The functionals associated to the different combination of overlap integrals  at the different densities are evaluated numerically from the simulations. This is obtained after the cooling/warming  procedures associated to the Pauli constraint.
From the average overlaps (on different microscopic configurations)  at different densities it is possible to obtain the linear system for the strength parameters whose solution will produce the reference EoS.

\section{Finite systems}
\subsection{Finite Range Interaction
 and reaction Mechanisms}
In this section  the above discussed effective interaction in CoMD calculations is used to study, as an example, the $^{64}Ni+^{48}Ca$ collision at different incident energies. This system  has been also widely experimentally investigated  (see  Ref. \cite{exp3} as an example).   
A comparison between reaction mechanisms produced using a ZR effective interaction ($m^{*}_{r}=1$) with parameter values modified for the CoMD  correlations (ZR-MD)   and the 
above-described finite range interaction (MDI-MD) will be illustrated in the next section.
In the same section we also add  the interesting comparison with the case of the MDI with parameters obtained in the MF scheme (MDI-MF). This comparison in fact allows to estimate how the typical
MD correlation affect the dynamics of the heavy collisions. 

\subsection{The interaction and the "ground state" configuration}
Moving on to finite systems the surface properties acquire a relevant role that cannot be described in the limit of an infinite NM.
Therefore a correction term $E_{s}$ is usually introduced in the 
total energy of A nucleons through the following expression
$E_{s}=\frac{C_{s}}{2}(4\pi \sigma_{r}^{2})^{1.5}\sum_{i=1,A}\nabla^{2}_{i}S^{i}_{v}$.
 
In CoMD for finite systems, this correction term is fixed, trough a warming-cooling procedure  applied to the microscopic configuration of the system under study and driven through the Pauli constraint. A stabilization stage follows.
In the present calculations, the process is stopped when the system reaches the given binding energy, the root-mean-radius, average kinetic energy, in a stable way (within 8\% of the requested values) for a  time interval of about 300 fm/c. 
For  the three mentioned interactions  MDI-MD, ZR-MD and MDI-MF good "ground state" properties are reached by setting  $C_{s}=-11$ MeV*fm$^{2}$,$C_{s}=3$ MeV*fm$^{2}$ and $C_{s}=-7$ MeV*fm$^{2}$ respectively.
The strength parameter of the effective interaction have been fixed to the values reported in the first row of Table 1, 
 Table 2 and Appendix B of Ref. \cite{npa}  for the cases
MDI-MF,MDI-MD and ZR-MD respectively.

\subsection{Selection and comparison of the reaction mechanisms and transverse flow}
Collisions in a wide range of impact parameters up to the "grazing" collision have been calculated with the CoMD model  
to identify different reaction mechanisms.
In Fig. 1  for the case ZR-MD the bi-dimensional plot $Z-V_{p}$
for the collision $^{64}Ni+^{48}Ca$ at different incident energy is shown.
Z represents the charge of the fragment formed after a maximum time of 350 fm/c.
$V_{p}$ represents the velocity of the produced fragments in the laboratory frame along the beam direction. 
Fragments  are identified with a minimum-spanning-tree method.
The figures refers to an impact parameter range
$b=0-10.5$ fm.
In order to testing the interactions, the present work will focus on the dynamics of processes producing 
fusion-incomplete fusion residues, fission and intermediate mass fragments in the mid-rapidity region in central/semi-central collisions.
The aforementioned process are well localized and almost exclusively selected in the
range of impact parameters $0-5$ fm
(the bumps corresponding to the bynary process are substantially absent).

In this case the fragments in the mid-rapidity region are mainly produced through the formation of one intermediate system in which the largest changes of density are produced.
Furthermore, the primary source is formed by target and projectile   nucleons having on average a  relative velocity  directly established through the beam energy (participant zone). Therefore we think that  the aforementioned processes are better suited to discuss the behaviours of the interaction  as a function of the incident energy. 
In particular in Fig. 1 the upper rectangles define the region where the fusion/incomplete fusion residues have been integrated.
Inside the lower rectangles will be evaluated the cross-section for production of Intermediate Mass Fragments (IMF) which are defined as fragments  having charge Z=3$\div$12. 
In the same region of velocity a kind of  "fission" of the hot residues is associated to the production, in the same event, of two fragments 1 and 2   heavier than the Nitrogen and having comparable charges (with the associate ratio of the charges Z1/Z2 greater than 0.8 and Z2 larger than or at most equal to Z1).
For the energies not shown in the figure the velocity ranges have been scaled proportionally to the center of mass (c.m.) velocity.
As observed before, in the present work the quantitative analysis will be focused on the impact parameter range $b=0-5$ fm where the above mechanisms are well localized.

\noindent
In the interval of considered impact parameters and for the energy interval 10-50 AMeV the mechanism is characterized by  the production of a bump in the $Z-V_{p}$  plots centred around the c.m. velocity.
In addition to the IMF production, heavy residues can be identified having charge Z higher than 20.
 This happens for both the MDI and the ZR cases cases.
\noindent
In these calculations due to the impact parameter selection the binary mechanisms are absent.
In Fig. 2 for the MDI-MD and ZR-MD cases, the integrated cross-section for the production of heavy residues (as  defined above) are shown as function of the incident energy.
They are represented by  black square and blue  
dots respectively.
In the inset panel the associated fission probabilities
estimated as: $P_{f}=\frac{N_{fis}}{N_{fus}+N_{fis}}$ are also plotted. In the above expression 
$N_{fis}$ and $N_{fus}$ indicate the selected number of events for fusion and fission respectively.
\begin{figure}
\centering
\includegraphics[width=8cm]{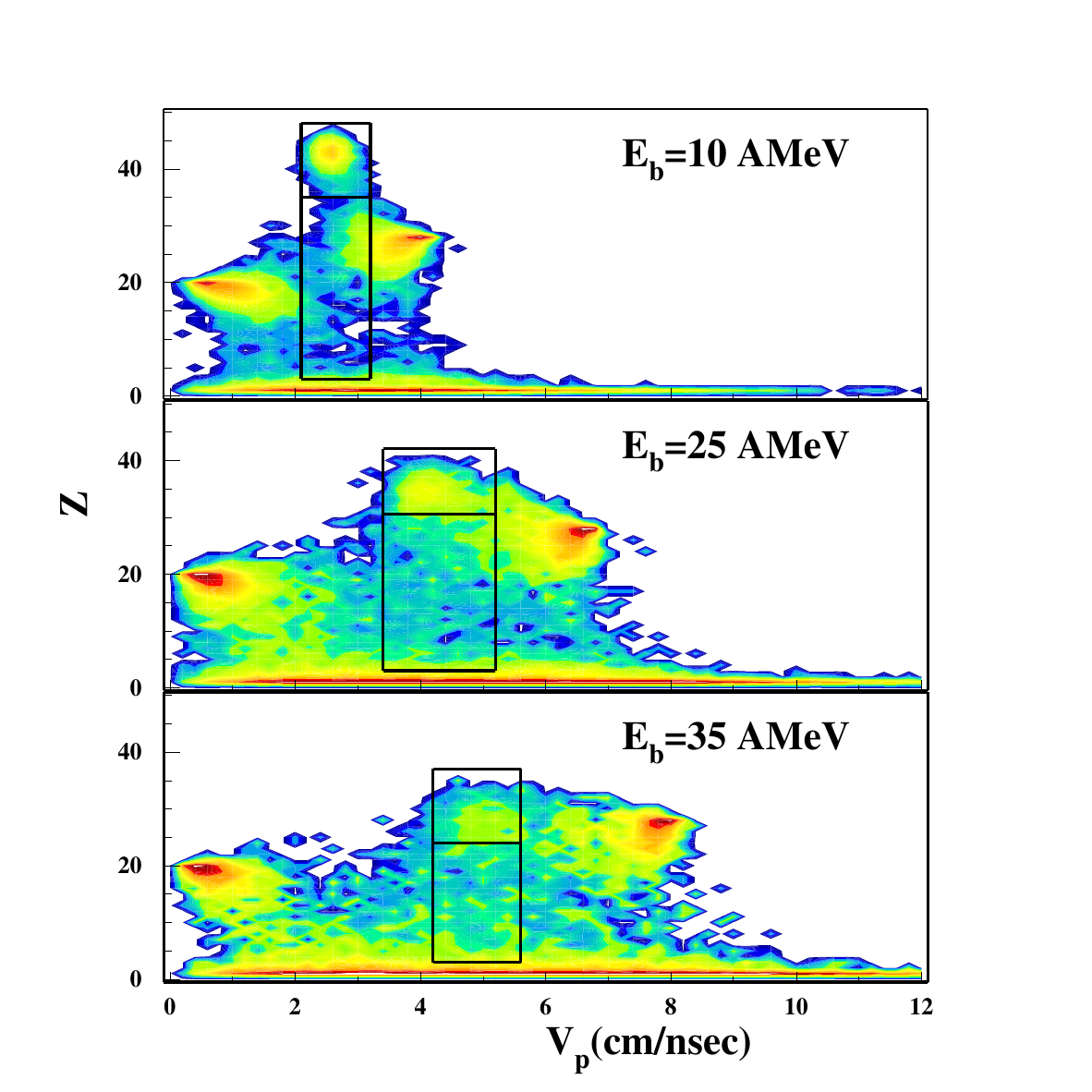}
         \caption{For the collisions 
$^{64}Ni+^{48}Ca$ (ZR-MD case), for the impact parameter range $b=0-10.5$ fm,  the
produced fragment charges Z versus the laboratory velocities $V_{p}$ along the beam direction are plotted for different incident energies.
The rectangles define the regions where the different reaction mechanisms are integrated (see text) (color).}
\end{figure}
\hspace{5mm}
\begin{figure}
\includegraphics[width=8.cm]{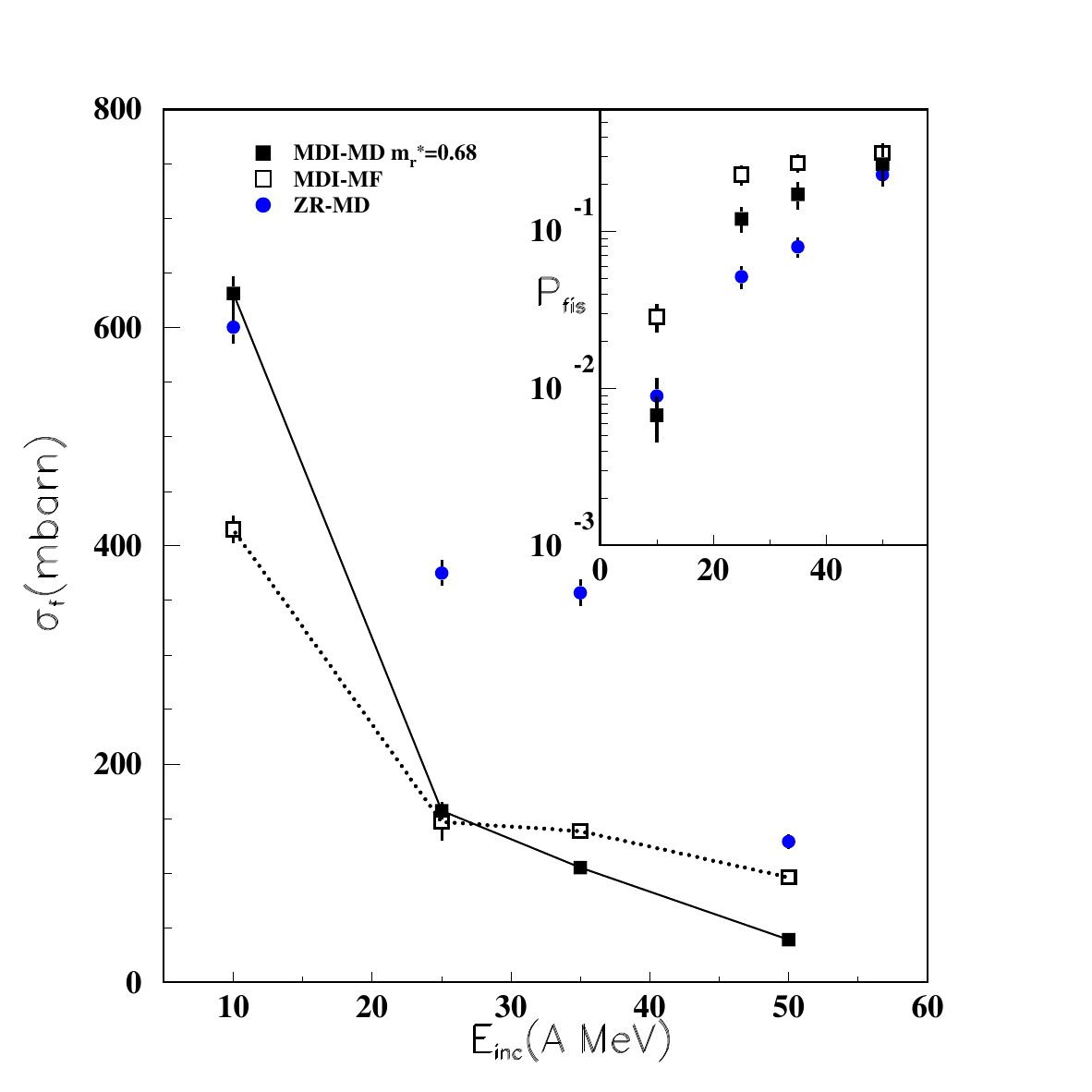}
         \caption{\label{Fig:3}  As a function of the incident energy $E_{inc}$ for the $^{64}Ni+^{48}Ca$ collision and
for an impact parameter range $b=0-5$ fm  the evaluated cross-section associated to the formation of an heavy residue is plotted.
The blue dots and the black square  symbols refer to the ZR-MD and MDI-MD cases respectively.
The open squares  represent the results for the MDI-MF case. 
In the inset the fission probabilities are also shown.
The lines joining the points MDI-MD and MDI-MF are plotted to simplify the comparison between the two cases. (color on-line)}
\end{figure}

\noindent

\noindent
Within this energy range, the globally more repulsive character of the finite range interaction case (MDI-MD)
gives rise to a more probable disassembly of the hot compound. It in fact produces a lower cross section for the 
formation of hot residues and an enhanced "fission" probability for  $E_{inc}$ at 25 and 35 AMeV. 
This more repulsive character of the MDI-MD is also confirmed through  the production of an higher rate of IMF. 
This is shown in Fig. 3 where the cross sections for producing at least one IMF in the mid-rapidity region and the related multiplicity are plotted in  panels a) and b).
We note that in both cases the rates of IMF have a maximum around 70 AMeV. The further
increase of the energy produces disassembly processes with an increasing
fraction of light particles with charge Z less than 3. 
This produces a lowering of the IMF rate in both  cases. 

\begin{figure}
\centering
\includegraphics[width=12.cm, height=8.5cm]{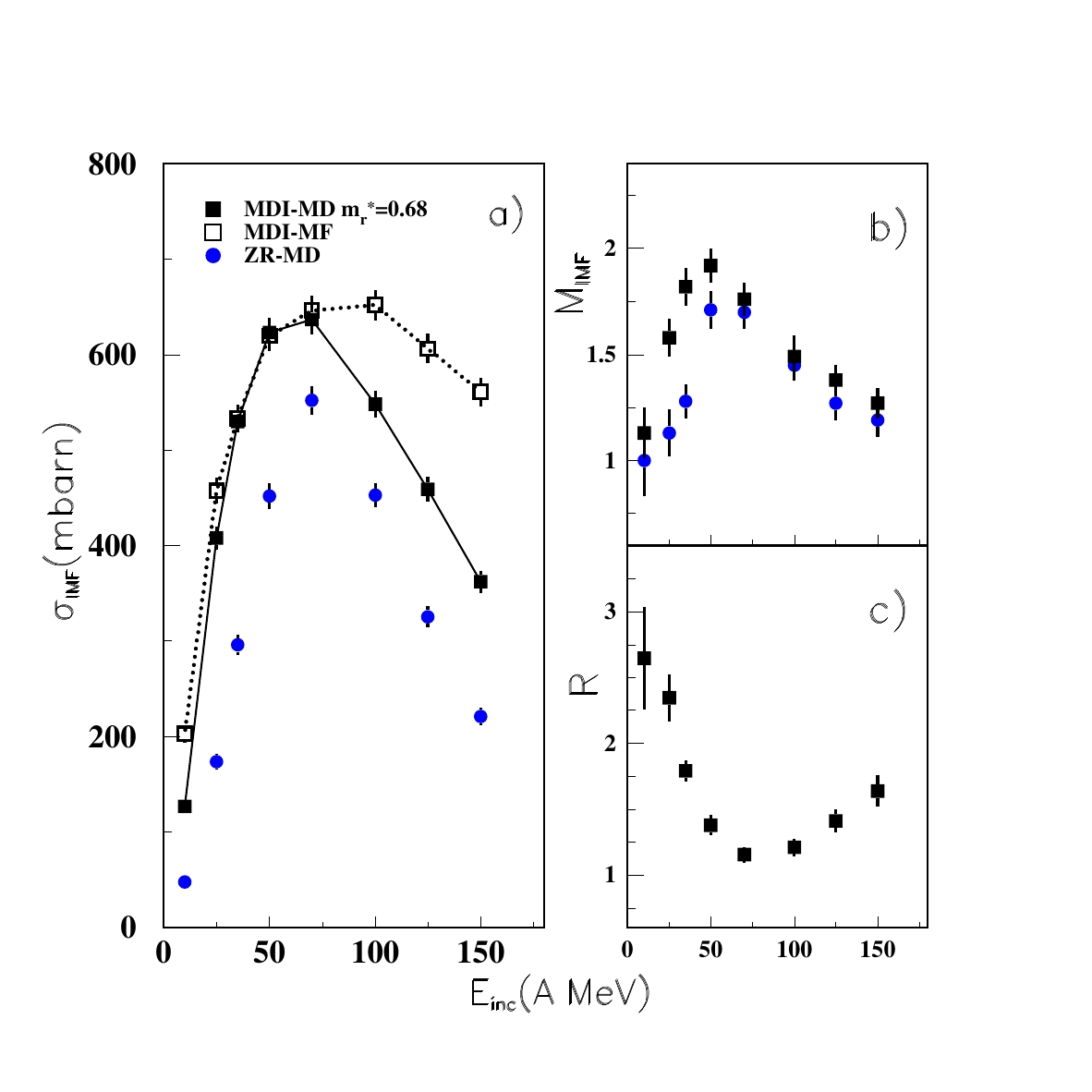}
         \caption{For the system $^{64}Ni+^{48}Ca$, in the impact parameters range  $b=0-5$ fm: panel a) shows for the MDI-MD, MDI-MF, ZR-MD  cases the 
cross section for the production of at least one IMF  as a function of the incident energy $E_{inc}$, panel b) shows the
associated multiplicity MDI-MD and ZR-MD , panel c) shows the ratio
$R=\sigma_{IMF}^{MDI-MD}/\sigma_{IMF}^{ZR-MD}$.
The lines joining the points MDI-MD and MDI-MF are plotted to simplify the comparison between the two cases. (color on-line)}
\end{figure}

\noindent
In both figures the open square symbols represent instead
the cross sections for the case MDI-MF.
The lines joining the points are plotted to simplify the comparison with the MDI-MD case. 
From this comparison it can be clearly observed how the corrections on the values of strength parameters
due to the discussed MD correlations can sensitively affect the studied quantities.

Going back to the earlier comparisons MDI-MD, ZR-MD,
in the panel c) it is shown the ratio 
$R=\sigma_{IMF}^{MDI-MD}/\sigma_{IMF}^{ZR-MD}$ associated to the IMF production. 
The ratio $R$ in this first part decreases  with the energy.

In the the analysed results at 100 AMeV and for the ZR case  a small bump at velocity  much higher than the c.m. one
appears in the $Z-V_{p}$ correlation plot (corresponding to $b\simeq 5$ fm). 
This hint of bump could be associated with the onset  of a precursor mechanism for the production of projectile-like fragments. This bump is
also present for the MDI-MD fm case but it is less pronounced.
Therefore, the presence of this other mechanism produces around the c.m. velocity a larger depletion in the $Z-V_{p}$ plot for the ZR-MD case with respect to the MDI-MD one and justifies the increasing behaviour of the ratio $R$ shown in Fig. 3c  from 100 up to 150 AMeV.

\noindent
The  global repulsive action associated to the 
finite range interaction strongly affects also the nucleon transverse flow as shown in Fig. 4. As an example for $b=3$ fm, it produces the disappearance of the balance energy (which it is seen for the ZR-MD case)  with positive slopes in  the explored energy range. 
In the upper energy limits the slopes for the two cases become comparable
being more affected by the nucleon-nucleon collision rate.

\noindent
Finally we observe that both the fast decreasing trend of the ratio R in the first 70 AMeV  (corresponding to about 35 AMeV of relative motion) and the behaviour of the transverse flow slopes as a functions of the beam energy could be interpreted trough the existence of a characteristic relative kinetic energy per nucleon   $E\simeq\frac{\hbar^{2}}{2m_{0}}\frac{2}{\xi^{2}}$  beyond which \textit{the momentum dependent effects related to the finite range, in the pre-equilibrium stage} of central/semi-central collisions  are attenuated. Here $\hbar\frac{\sqrt{2}}{\xi}$ establish the characteristic width of the MDI.

\begin{figure}
\centering

\includegraphics[width=12.cm, height=10.cm]{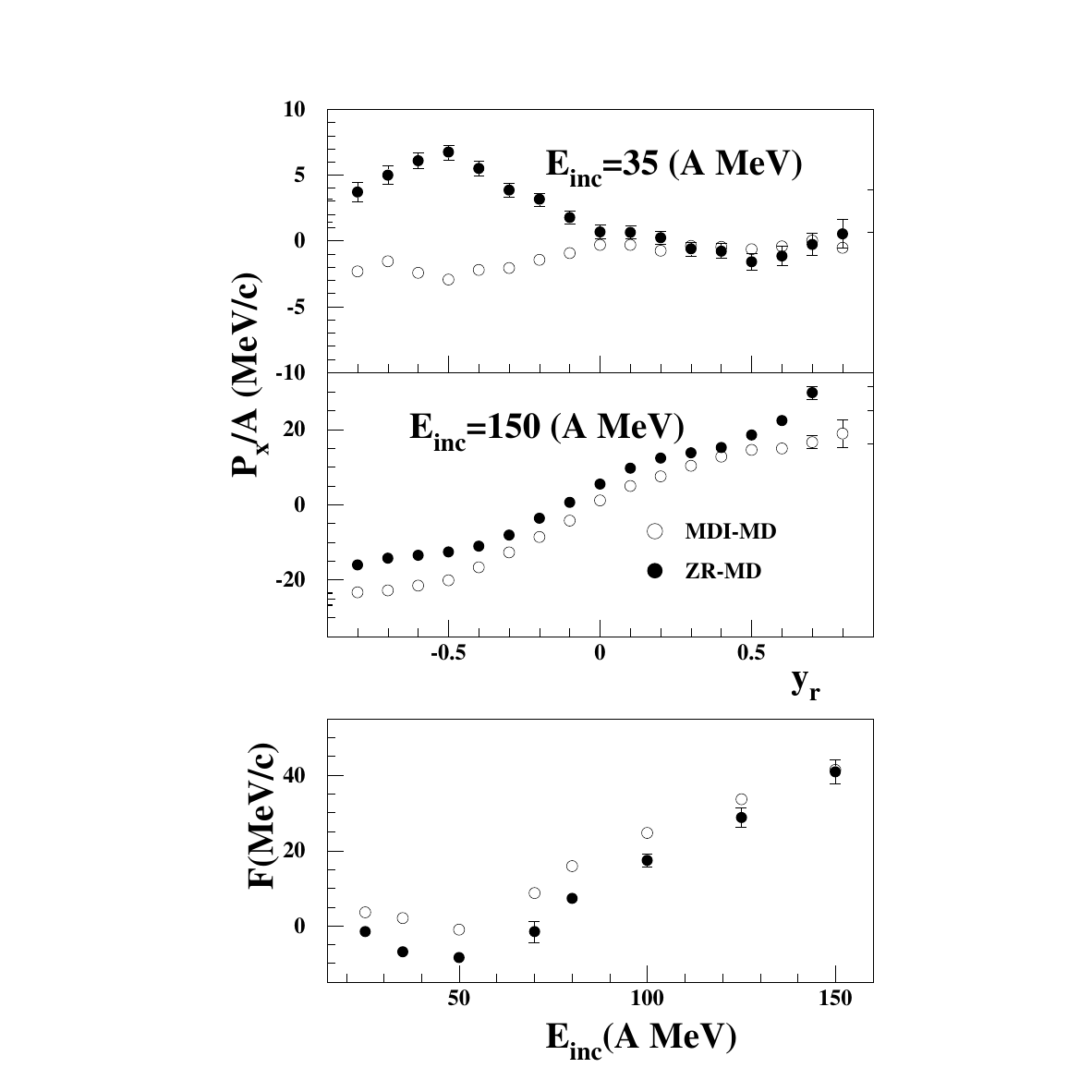}
         \caption{\label{Fig:7}  In the two upper panels, for the zero and the ﬁnite range interaction (MDI)  different values of the nucleon effective mass the average transverse momentum per nucleon ($b=3$ fm) is  plotted  as a function of the c.m. reduced rapidity.
In the bottom panel the slope parameters as  a function of the incident energy are also shown.}

\end{figure}

\subsection{Conclusive remark}
In this work, starting from some reference EoS properties at the saturation density, the effects of the finite range interaction as compared to  the case of a zero range interaction for the system  $^{64}Ni+^{48}Ca$ system has been studied
in semi-central collision at different energies by means of the CoMD model. 
Furthermore, for the same system  the effects of the many-body correlations in the EoS obtained in of the model (as compared to a MF approach) in  fusion-like cross-section fission probability and IMF production
in central collisions  were found to be non-negligible.  
Finally, it's worth noting  that
even if the obtained numerical results are strictly valid for the specific model calculations,
the rather general feature of the discussed correlations that are associated to the wave-packets dynamics and to the Pauli principle could give a wider meaning to the 
the comparisons illustrated in this work.


\end{document}